\documentclass[aps,prb,twocolumn,superscriptaddress]{revtex4-1}

\usepackage{graphicx}
\usepackage{color}
\usepackage{here}

\begin{document}







\title{Determination of the local structure of Sr$_{2-x}$M$_x$IrO$_4$ (M = K, La) as a function of doping and temperature} 

\author{K. Terashima}
\email[]{k-terashima@cc.okayama-u.ac.jp}
\affiliation{Research Institute for Interdisciplinary Science, Okayama University, Okayama, 700-8530, Japan}

\author{E. Paris}
\altaffiliation{Research Department Synchrotron Radiation and Nanotechnology, Paul Scherrer Institut, CH-5232 Villigen PSI, Switzerland}
\affiliation{Dipartimento di Fisica, Universit\'{a} di Roma ``La Sapienza'' - P. le Aldo Moro 2, 00185, Roma, Italy}

\author{E. Salas-Colera}
\affiliation{Instituto de Ciencia de Materiales de Madrid, ICMM-CSIC, Sor Juana In\'es de la Cruz 3, 28049 Madrid, Spain}
\affiliation{Spanish CRG BM25 Spline, ESRF - The European Synchrotron, 71 avenue des Martyrs, 38043 Grenoble, France}

\author{L. Simonelli}
\affiliation{CELLS - ALBA Synchrotron Radiation Facility, Carrer de la Llum 2-26, 08290, Cerdanyola del Valles, Barcelona, Spain}

\author{B. Joseph}
\affiliation{ELETTRA, Sincrotrone Trieste, Strada Statale 14, Km 163.5, Basovizza, 34149 Trieste, Italy}

\author{T. Wakita}
\affiliation{Research Institute for Interdisciplinary Science, Okayama University, Okayama, 700-8530, Japan}

\author{K. Horigane}
\affiliation{Research Institute for Interdisciplinary Science, Okayama University, Okayama, 700-8530, Japan}

\author{M. Fujii}
\affiliation{Department of Physics, Okayama University, Okayama, 700-8530, Japan}

\author{K. Kobayashi}
\affiliation{Research Institute for Interdisciplinary Science, Okayama University, Okayama, 700-8530, Japan}
\affiliation{Department of Physics, Okayama University, Okayama, 700-8530, Japan}

\author{R. Horie}
\affiliation{Research Institute for Interdisciplinary Science, Okayama University, Okayama, 700-8530, Japan}

\author{J. Akimitsu}
\affiliation{Research Institute for Interdisciplinary Science, Okayama University, Okayama, 700-8530, Japan}

\author{Y. Muraoka}
\affiliation{Research Institute for Interdisciplinary Science, Okayama University, Okayama, 700-8530, Japan}
\affiliation{Department of Physics, Okayama University, Okayama, 700-8530, Japan}

\author{T. Yokoya}
\affiliation{Research Institute for Interdisciplinary Science, Okayama University, Okayama, 700-8530, Japan}
\affiliation{Department of Physics, Okayama University, Okayama, 700-8530, Japan}

\author{N. L. Saini}
\affiliation{Dipartimento di Fisica, Universit\'{a} di Roma ``La Sapienza'' - P. le Aldo Moro 2, 00185, Roma, Italy}

\begin{abstract}
The local structure of correlated spin-orbit insulator Sr$_{2-x}$M$_x$IrO$_4$ (M = K, La) has been investigated  by Ir L$_3$-edge extended x-ray absorption fine structure measurements.  The measurements were performed as a function of temperature for different dopings induced by substitution of Sr with La or K.  It is found that Ir-O bonds have strong covalency and they hardly show any change across the N\'eel temperature.  In the studied doping range, neither Ir-O bonds nor their dynamics, measured by their mean square relative displacements, show any appreciable change upon carrier doping, indicating possibility of a nanoscale phase separation in the doped system.  On the other hand, there is a large increase of the static disorder in Ir-Sr correlation, larger for K doping than La doping.  Similarities and differences with respect to the local lattice displacements in cuprates are briefly discussed. \\
\end{abstract}







\maketitle

\section{Introduction}
	
	Transition-metal oxides \cite{RMP} have been one of the major research subjects in condensed matter physics for last few decades, stimulated by large electron-electron correlation in 3{\it d}-electron systems.  Recently, 5{\it d}-electron systems are acquiring enormous attention in the field since the spin-orbit interaction energy in these is comparable to the Coulomb interaction and transfer integral energy scale and hence novel phenomena are expected to emerge due to their interplay \cite{Jackeli}.  
	
	Among 5{\it d}-electron systems, Sr$_2$IrO$_4$ is an insulator with an antiferromagnetic order at {\it T}$<$240 K \cite{Crawford}.  Recently, the electronic structure of Sr$_2$IrO$_4$ has been reported to be well described by $J_{eff}$ = 1/2 ground state \cite{KimPRL}, and its insulating behavior has been realized due to the splitting of $J_{eff}$ = 1/2 band into lower and upper Hubbard band.  This material has been regarded as an analogue of 214-type cuprate superconductors ({\it s} = 1/2) in several aspects, such as K$_2$NiF$_4$-type crystal structure \cite{Crawford}, antiferromagnetic magnetic ordering, and correlated insulating behavior.  In the crystal structure, the Ir-O octahedra is elongated along {\it c}-axis and further rotated by $\sim$11 degree around {\it c}-axis and the material forms a canted antiferromagnetic ordering \cite{KimScience}.  Although there has been no reports on the temperature dependence of local structure, an x-ray diffraction study has argued that the temperature dependence of resistivity may be related with atomic coordinates of Ir and O atoms \cite{Bhatti}.
	
	For the carrier-doped Sr$_2$IrO$_4$ system, the emergence of superconductivity has been predicted theoretically in the context of the analogy with cuprates \cite{FWang}. Although superconductivity in this class of material has not been reported to date, angle-resolved photoemission studies have reported that the electronic structure in lightly carrier-doped samples are similar with cuprates in several aspects. Furthermore, the existance of anisotropic excitation gap reminiscent of pseudogap have been reported \cite{KimNP, delaTorre, Cao, KT}.
	
	In doped Mott insulators such as the cuprates, not only the spin but also the lattice degree of freedom have been investigated, leading to the conclusion that the strong interaction among phonons, spin fluctuations, and the electronic structure should play a role for the emergence of high-{\it T}$_c$ superconductivity \cite{Keimerreview, Lanzara, Gweon}.  It has been also found that the charge/spin density wave is likely to commonly exist in the pseudogapped underdoped region of cuprates \cite{Tabis}.  Among the lattice-sensitive experimental methods, extended x-ray absorption fine structure (EXAFS) is a unique technique to resolve local atomic displacements that turned out to be sensitive to the occurance of a charge ordering in cuprates \cite{Saini1, Saini2}.  On the other hand, the corresponding work of temperature-dependent local lattice displacements has not been reported in Sr$_2$IrO$_4$ series although there are several signatures of the pseudogap \cite{KimNP, delaTorre, Cao, KT} in the system.  Furthermore, the bilayer compound Sr$_3$Ir$_2$O$_7$ with Ruddlesen-Popper structure has been found to show an indication of a possible charge order \cite{ChuNM}.  In Sr$_2$Ir$_{1-x}$Rh$_x$O$_4$ also, optical second harmonic generation and neutron diffraction measurements have indicated a symmetry lowering below a certain temperature due to an emergence of an ordered phase, where both the spatial inversion and rotational symmetries of the tetragonal lattice are broken \cite{ZhaoSHG, Jeong}.  This state has been interpreted as a loop-current ordered state that was initially proposed as an explanation for the pseudogap state in cuprates.  Possible existence of a spin density wave (SDW) state has been also reported \cite{XChen}.  Despite these similarities with cuprates, there have been no report of superconductivity yet in Sr$_2$IrO$_4$ upon doping.  Therefore, a further systematic study on how the lattice degree of freedom reacts upon doping in the system should provide important information in clarifying the metallization process of correlated spin-orbit insulator Sr$_2$IrO$_4$.

	In this study, we report doping- as well as temperature-dependence of the local structure around Ir atom in both hole- and electron-doped Sr$_{2-x}$M$_x$IrO$_4$ (M = K, La), aiming to find out possible coupling between local lattice degrees of freedom with the ordering of spin including antiferromagnetism.  The work is also aimed to find possible signature of ordering involving local lattice like charge density wave as has been seen in cuprates.  Here, the carriers were introduced by a partial substitution for Sr atom in the Sr-O block layer for both type of doping.  By a systematic study, we have found that the in-plane Ir-O bond has a strong covalency resulting in an Einstein-frequency as high as 800 K, and the bond is hardly affected by carrier doping nor the occurance of magnetic order in the system as a function of temperature.  We did not find marked signature of formation of ordered state in the lattice response within experimental uncertainties, while we do see a substantial change in the Ir-Sr correlations, affected more by hole-doping than electron-doping.  The observed local lattice response against doping differs from that of cuprates, indicating that a manipulation of hard local mode of in-plane Ir-O bond may have some key role for possible superconductivity in these materials.

\section{Experimental}
	Polycrystalline samples of Sr$_{2-x}$M$_x$IrO$_4$ (M = La, K) were prepared by conventional solid-state reactions.  A mixture of SrCO$_3$, K$_2$CO$_3$, La$_2$O$_3$ and IrO$_2$ was ground and further mixed by planetary ball-milling (Fritsch, P-7) at a rotation rate of 400 rpm for 3 h with 15 (5 mm-diameter) and 10 (10 mm-diameter) ZrO$_2$ balls.  The resulting powders were calcined in air at 1150 $^{\circ}$C for 15 min \cite{Horigane}.  Nominal values of substituting atoms of La for Sr$_{2-x}$La$_x$IrO$_4$ was 0.075. The amount of K atoms in Sr$_{2-x}$K$_x$IrO$_4$ samples were determined to be {\it x} = 0.04 and 0.055 by energy dispersive x-ray spectrometry measurements. Hereafter we call those samples as parent, La, K0.04, K0.055.   Phase purity of samples were examined by x-ray diffraction measurements (shown in supplemental information).  N\'eel temperature ({\it T}$_N$) were evaluated for all the samples by measuring their magnetization curve, and obtained values are 240 K, 200 K, 235 K, and 225 K for parent, La, K0.04, and K0.055, respectively (see also supplemental information for magnetization curve and resistivity). Judging from the amount of dopant atom and {\it T}$_N$ value, the doping amount of La- and K-doped sample of the present study would correspond to the samples where excitation-gapped state has been observed by ARPES \cite{Cao, KT}, and the La amount is close to the sample where a signature of SDW state has been observed \cite{XChen}. The hole doping amount of our K-doped samples would also be in the regime where a symmetry-broken state has been observed in hole-doped Sr$_2$Ir$_{1-x}$Rh$_x$O$_4$ \cite{ZhaoSHG, Jeong}.
	
	Ir L$_3$-edge ({\it E}$\sim$11 keV) x-ray absorption measurements were carried out at the Spline beamline \cite{Spline} of the European Synchrotron Radiation Facility where Si(111) double crystal monochromator was used to obtain the energy resolution $\Delta$$E/E$ of 1.4$\times$10$^{-4}$.  The powder samples were mixed with cellulose matrix and then pelletized for transmission measurements, to reach the desired thickness for the absorption jump to be $\sim$1 at the Ir L$_3$-edge.  Fluorescence signals of samples and Pt L$_3$ edge of Pt film for energy reference placed at down stream of the beam were recorded simultaneously (not shown).  Several absorption scans were acquired at each temperature to ensure spectral reproducibility for each sample and to estimate the statistical error. Both of the statistical error and the error coming from the correlation between parameters were taken into account for the error bars of obtained physical parameters.

\begin{figure}[h]
\includegraphics[width=8.3cm]{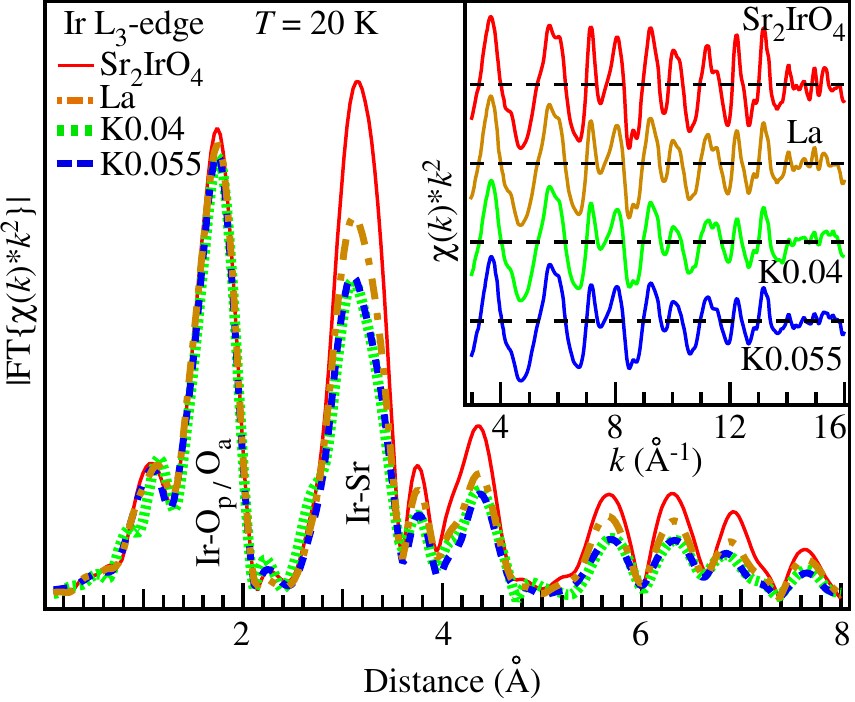}
\caption{Fourier transform magnitudes of Ir L$_3$-edge EXAFS ({\it k$^2$}-weighted) measured on Sr$_2$IrO$_4$ (solid line), Sr$_{1.925}$La$_{0.075}$IrO$_4$ (dotted dashed line), Sr$_{1.96}$K$_{0.04}$IrO$_4$ (dotted line), and Sr$_{1.945}$K$_{0.055}$IrO$_4$ (dashed line) samples at 20 K. Here, the data are not corrected by the phase shifts.  Inset shows the corresponding EXAFS oscillations of $\chi(k)*k^2$.}
\end{figure}

\section{Results and Discussion}

\begin{figure*}[t]
\includegraphics[width=18cm]{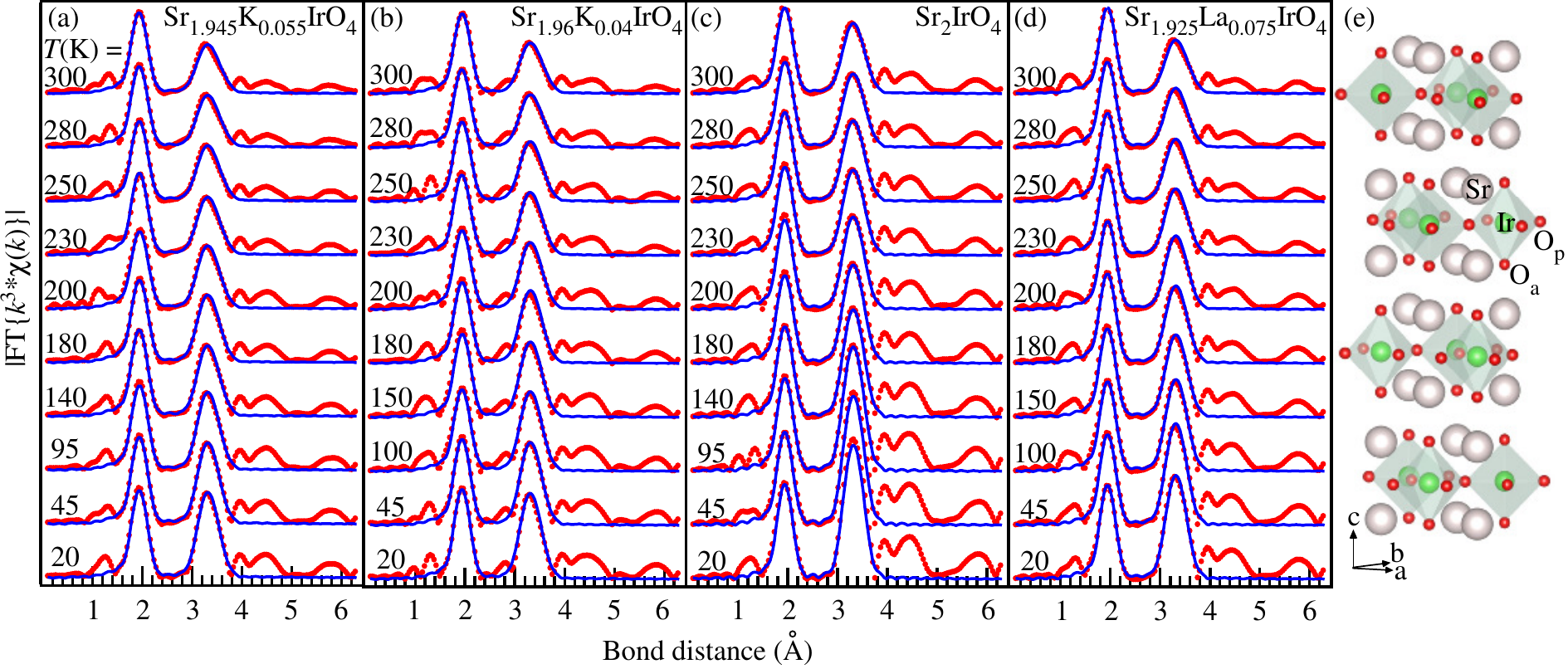}
\caption{Temperature dependent Fourier transform magnitudes of Ir L$_3$-edge EXAFS for  Sr$_{1.945}$K$_{0.055}$IrO$_4$ (a), Sr$_{1.96}$K$_{0.04}$IrO$_4$ (b), Sr$_2$IrO$_4$(c), and Sr$_{1.925}$La$_{0.075}$IrO$_4$ (d) after phase shift correction.  Red dots represent experimental data, while blue lines show the model fits considering three shells namely Ir-O$_p$ and Ir-O$_a$, and Ir-Sr. (e) Structural image of Sr$_2$IrO$_4$.}
\end{figure*}	

	Figure 1 shows the magnitude of Fourier transforms of Ir L$_3$-edge EXAFS oscillations extracted from the x-ray absorption spectra on parent Sr$_2$IrO$_4$ and La, K0.04, and K0.055 samples at {\it T} = 20 K.  The Fourier transforms were obtained using a Gaussian window ({\it k}-range 3.1-16 \AA$^{-1}$) and are not corrected for the phase shifts thus represent raw data.  The first peak appears around 1.8 \AA$ $ corresponds to the bond distances \cite{Ranjbar} between Ir and in-plane oxygen (Ir-O$_{p}$, $\sim$1.98 \AA$ $) and Ir and apical oxygen (Ir-O$_{a}$, $\sim$2.05 \AA$ $) atoms, while the second peak at around 3 \AA $ $ corresponds to the bond distance between Ir and Sr atoms ($\sim$3.35 \AA$ $).  The relatively small peak located at around 3.8 \AA$ $ corresponds to Ir-O-Ir multiple scattering, overlapped with distant atom contributions.  The first peak (Ir-O) is less affected than other peaks upon Sr substition with La or K, which is expected since we substitute atoms at Sr site.   Inset shows $k^2$-weighted $\chi(k)$ EXAFS oscillations for each sample at the same temperature.  All spectra show overall correspondence each other and signals tend to be damped at higher {\it k}-region beyond $\sim$14 \AA$^{-1}$.  Next we take into account the phase shift correction to quantify local atomic displacements.

	Figure 2 shows the temperature dependent Fourier transforms of $k^3$-weighted Ir L$_3$-edge EXAFS after phase shift correction.  Spectra in (a)-(d) are shown in the same vertical scale with the same amount of offset for a realistic comparison.  It is clear from the figure that with increasing temperature, the spectral intensity $\sim$3.5 \AA$ $ is more suppressed than that of $\sim$2 \AA$ $, reflecting the different bond strength characteristics of Ir-O and Ir-Sr distances. In addition, judging from the manner of the suppression in the spectral contribution from distant atom contributions ($>$4 \AA$ $), it can be said that K atom introduces larger disorder than La atom as a substituent.  For a realistic comparison of the local structure of different compounds, we have evaluated the local structure parameters by standard EXAFS model fits using single scattering approximation \cite{Bunker}.  Namely, the bond distances and the mean square relative displacement (MSRD) of the absorber-backscatterer pair of atoms are obtained. The EXCURVE 9.275 code \cite{Excurve} was used for the model fits.  The model fits are displayed by solid lines in Fig. 2.  As starting values for the fit, we have used the structural parameters reported at room temperature for the parent compound \cite{Ranjbar}.  The present analysis is modelled by three shells, namely Ir-O$_{p}$ and Ir-O$_{a}$, and Ir-Sr, where the photoelectron phase shifts for each bond is taken into account.  In the model fit, the passive electrons reduction factor S$_0$$^2$ is set to 0.95. The number of neighboring atoms N$_i$ are 4 for Ir-O$_p$, 2 for Ir-O$_a$, and 8 for Ir-Sr.  The photoelectron energy zero ({\it E}$_0$) was fixed after fit trials on different scans.  Six parameters, namely the bond distances and MSRDs of Ir-O$_p$, Ir-O$_a$, and Ir-Sr bonds are varied in the model fit.  The $R$-range for the model fits is 1.5-4.0 ($\Delta R$ = 2.5) while the $k$-range is 3.1-16.0 ($\Delta k$ = 12.9) with the number of independent parameters (2$\Delta R\Delta k$/$\pi$) being about 20. In the model fit, an attempt was made to include two different Ir-O$_a$ bonds as suggested by a second harmonic light reflection measurement \cite{Torchinsky}, however, we did not find any evidence of such bond distances in our studied system.

	Figure 3 shows temperature dependence of bond lengths of Ir-O$_{p}$, Ir-O$_{a}$, and Ir-Sr in K0.055, K0.04, parent, La samples, obtained by the EXAFS model fits.  The MSRDs of each bond are shown in Fig. 4.  The dashed lines in Fig. 4 denote the fit result of MSRD by the correlated Einstein-model \cite{Sevillano}; $\sigma^2$ = $\sigma_0^2$ + $\sigma^2({\it T})$ where $\sigma_0^2$ denotes static part and $\sigma^2({\it T})$ stands for the dynamic part.  The vertical dotted lines in Fig. 3 and 4 indicate {\it T}$_N$ estimated from magnetization measurments for all the samples.   We found that the local structural parameters determined in the present study for the parent compound are consistent with those reported by neutron diffraction experiments \cite{Crawford, Huang, Shimura, Ye} (see also supplemental information for the comparison with the structural parameters reported earlier). We have also found that within the experimental uncertainties, neither local bond distance nor MSRD show any evident response against the formation of magnetic ordering in all the samples.  The mean value of Ir-O$_p$ bond distance tend to decrease by K-substitution and increase by La-substitution.  On the contrary, the mean value of Ir-O$_a$ bond distance tend to increase by K-substitution and decrease by La-substitution.  These tendencies are consistent with the earlier XRD report on Ba$_{2-x}$M$_x$IrO$_4$ (M = K, La) \cite{Okabe}. Regarding the temperature dependence of bond length, it is commonly observed in all the samples that the thermal expansion of Ir-O$_{p}$ bond seems to be smaller than that of Ir-O$_{a}$ bond, indicating the strong covalency between Ir and planar O atom.  The presence of strong covalency in Ir-O bond is also consistent with almost temperature-independent behavior of MSRD in Fig. 4 and estimated Einstein temperature is more than 800 K.  Interestingly, the strong covalency and high Einstein temperature of Ir-O$_{p}$ bond is not altered by the nature of carrier doping (electron and hole). $\sigma_0^2$ of Ir-O$_p$ remained as small as $\sim$0.001 \AA$^2$$ $ for all the samples.  Ir-O$_{a}$ bond seems to have slightly higher $\sigma_0^2$ value than Ir-O$_p$ by $\sim$0.001 \AA$^2$ $ $ with similarly high Einstein temperature, and the bond remained also unaffected by carrier doping although it should be mentioned that the experimental uncertainty of Ir-O$_a$ is larger than that of Ir-O$_p$ due to the proximity of those two bond lengths.  On the other hand, we have found notable increase of MSRD value of Ir-Sr bond in doped samples, that is mainly attributed to the increase of static part ($\sigma_0^2$).  The dynamical part ($\sigma^2({\it T})$) of Ir-Sr bond is much less affected resulting in almost unchanged Einstein temperatures in doped samples ($\sim$240 K for all).   Judging from $\sigma_0^2$ values, the degree of static disorder remained the same in two K-doped samples, although {\it T}$_N$ values and K amounts are different.  There is a tendency that the increase in static disorder at K-doped samples are higher than that at La-doped sample, which can be attributed to the larger difference of ionic radius between Sr$^{2+}$ (1.31 \AA$ $) and substituent atoms, namely ionic radius of K$^+$ is 1.55 \AA$ $  while that of La$^{3+}$ is 1.216 \AA$ $ in 9-coodination \cite{Shannon}.  

\begin{figure}[t]
    \includegraphics[width=8.3cm]{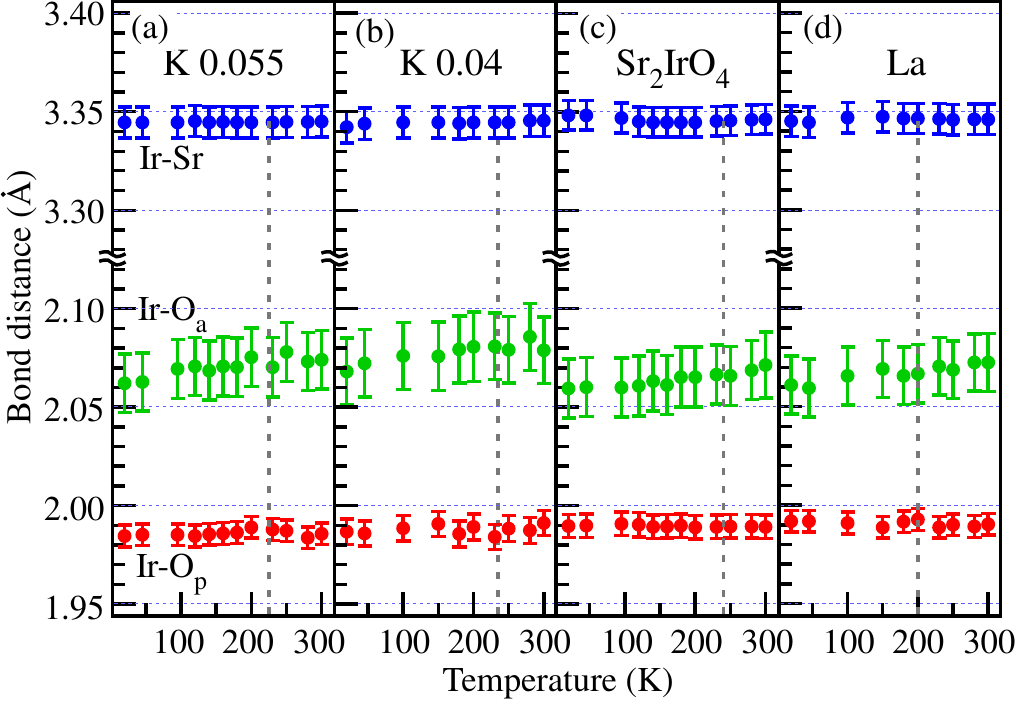}
\caption{Temperature dependence of bond length around Ir atom derived from model fit shown in Fig. 2. Vertical dotted lines denote {\it T}$_N$ for each sample.}
\end{figure}		

\begin{figure}[t]
\includegraphics[width=8.3cm]{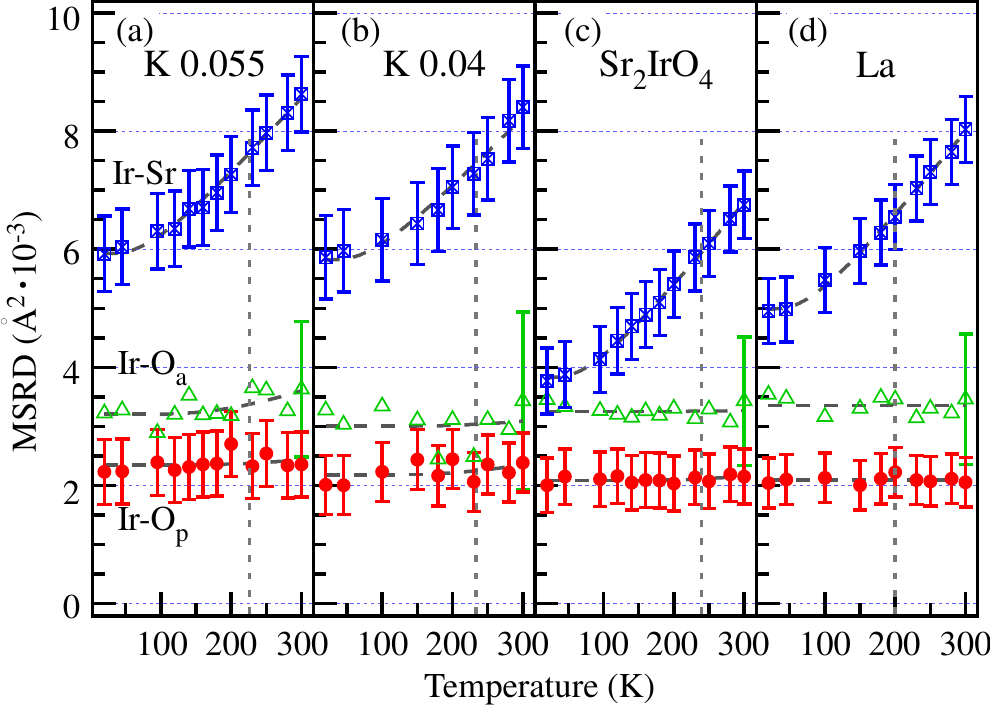}
\caption{Mean square relative displacement of Ir-O$_{p}$, Ir-O$_{a}$, and Ir-Sr bond distance as a function of temperature.  Dashed lines show the Einstein model fit for each bond.  For Ir-O$_a$, typical magnitude of error is put only for 300 K data for better visualization of data, avoiding error bars to highly overlap each other and mask other data. Vertical dotted lines denote {\it T}$_N$ for each sample.}
\end{figure}

	Let us attempt to understand possible implications of the present findings.  Our current observation of negligible softening in Ir-O$_{p}$ bond with high Einstein temperature suggests that the high covalency in Ir-O may not permit charge doping in the IrO$_2$ sublattice to obtain mobile carriers, as it is discussed to happen in nickelate heterostructures \cite{Grisolia}.  Therefore the carriers doped by the substitution in Sr-O layer would tend to localize, which may cause a nano-scale spatial phase separation.  In fact, there are scanning tunneling spectroscopy studies reporting a signature of phase separation between large-gap (Mott gap) state and small-gap state \cite{Chen, Battisti}.  Recent ARPES studies also report the coexistence of lower Hubbard band and in-gap band, which can be also reconciled by the possible phase separation \cite{KT, Brouet}.  In case of cuprates, the in-plane Cu-O bond tend to be softer \cite{Saini2}, allowing the doped charges to be mobile in CuO$_2$ plane.  It should be noted that cuprates are strongly affected by the Jahn-Teller distortion that can locally relax with doping, while such a distortion of IrO$_6$ octahedra is much less in Sr$_2$IrO$_4$.  We also note that in case of sister compound Sr$_3$Ir$_2$O$_7$, the system also shows phase separation at lightly doped regime but becomes metal \cite{Hogan} when La is doped more than 4 \%, while Sr$_2$IrO$_4$ does not in a similar doping level.  Such a difference can be caused by the difference of the band width and the dimensionality of the parent material \cite{Yamasaki}.  Apart from such other factors, a control of covalency in in-plane Ir and O atoms may be important to induce superconductivity in Sr$_2$IrO$_4$ system.

It is worth recalling that the formation of excitation-gapped state has been observed in the electronic structure of both hole- and electron-doped systems by ARPES \cite{delaTorre, Cao, KT}.  The existence of symmetry-broken state has been observed by second harmonic light reflection \cite{ZhaoSHG} and neutron diffraction \cite{Jeong} techniques in hole-doped Sr$_2$Ir$_{1-x}$Rh$_x$O$_4$, where this state has been discussed in relation with the excitation-gapped state \cite{ZhaoSHG}. In addition, recent magnetic resonant x-ray scattering measurement on Sr$_{2-x}$La$_x$IrO$_4$ has suggested the presence of SDW state \cite{XChen}.  In the current EXAFS study of Ir-O and Ir-Sr local lattice response, we did not observe any significant influence of the formation of such or any other ordered state in the temperature dependence within available doping range.  This is in constrast to the cuprate case, where the formation of a charge-ordered state triggers a characteristic softening in the in-plane Cu-O bond \cite{Saini1, Saini2} as revealed by polarized EXAFS.  Although no indication of such response of the local structure has been seen in the current study with certain experimental uncertainty (enhanced by the proximity of Ir-O$_p$ and Ir-O$_a$ bond lengths), further detailed study of polarized EXAFS using single crystals as well as in wider doping range should be helpful to obtain bond-resolved information on the local structure.

\section{Conclusions}

	In summary, doping- and temperature-dependence of atomic displacements around Ir atom are studied by Ir L$_3$ EXAFS in Sr$_{2-x}$M$_x$IrO$_4$ (M = K, La).  We have found that Ir atoms form strong covalent bond with O atoms with high Einstein-frequency, which does not get altered by electron or hole doping.  The configurational disorder in Ir-O layer seems to remain unaffected by doping while it is different for Ir-Sr.  The former observation is in contrast to the cuprate case, where doped charge induces softening in in-plane lattice displacement, and this difference may be one of the intervening factors for the occurance of superconductivity in Sr$_2$IrO$_4$ system.


\begin{acknowledgments}
We thank ESRF staff for support in the EXAFS data collection.  K. T. and T. W. would like to acknowledge the hospitality at the Sapienza University of Rome.  This research was partially supported by the Program for Promoting the Enhancement of Research University from MEXT, the Program for Advancing Strategic International Networks to Accelerate the Circulation of Talented Researchers from JSPS (R2705), and JSPS KAKENHI (Grants Nos. 2704, 25000003, 26247057, and 15H05886).  This work is a part of the executive protocol of the general agreement for cooperation between the Sapienza University of Rome and Okayama University, Japan.
\end{acknowledgments}






\clearpage
\widetext
\begin{center}
\textbf{\large Supplemental Information}
\end{center}

\setcounter{figure}{0}
\setcounter{section}{0}
\setcounter{page}{1}
\renewcommand{\thefigure}{S\arabic{figure}}

\section{Sample properties}
	Figure S1(a) show powder x-ray diffraction patterns of  Sr$_{2-x}$M$_x$IrO$_4$ (M = La, K, {\it x} = 0.075 for La and 0.055 for K), measured using a conventional x-ray spectrometer with a graphite monochromator (RINT-1100, Rigaku).  The parent sample showed tiny impurity peak (2$\theta$$\sim$18 deg) while no such intensity was observed in La and K0.055 samples, indicating high phase purity of samples. Figs. S1(b) and (c) show the temperature dependence of the resistivity and the magnetic susceptibility of samples. The electrical resistivity was measured by a conventional dc four-probe method. The magnitude of the resistivity was reduced by both type of doping but samples remained insulating.  Magnetic susceptibility measurements were performed using a superconducting quantum interference device magnetometer (Quantum Design MPMS-R2).  Both of the N\'eel temperature and the magnetic moment of samples were reduced by La or K substitution for Sr.

\begin{figure}[h]
\includegraphics[width=16 cm]{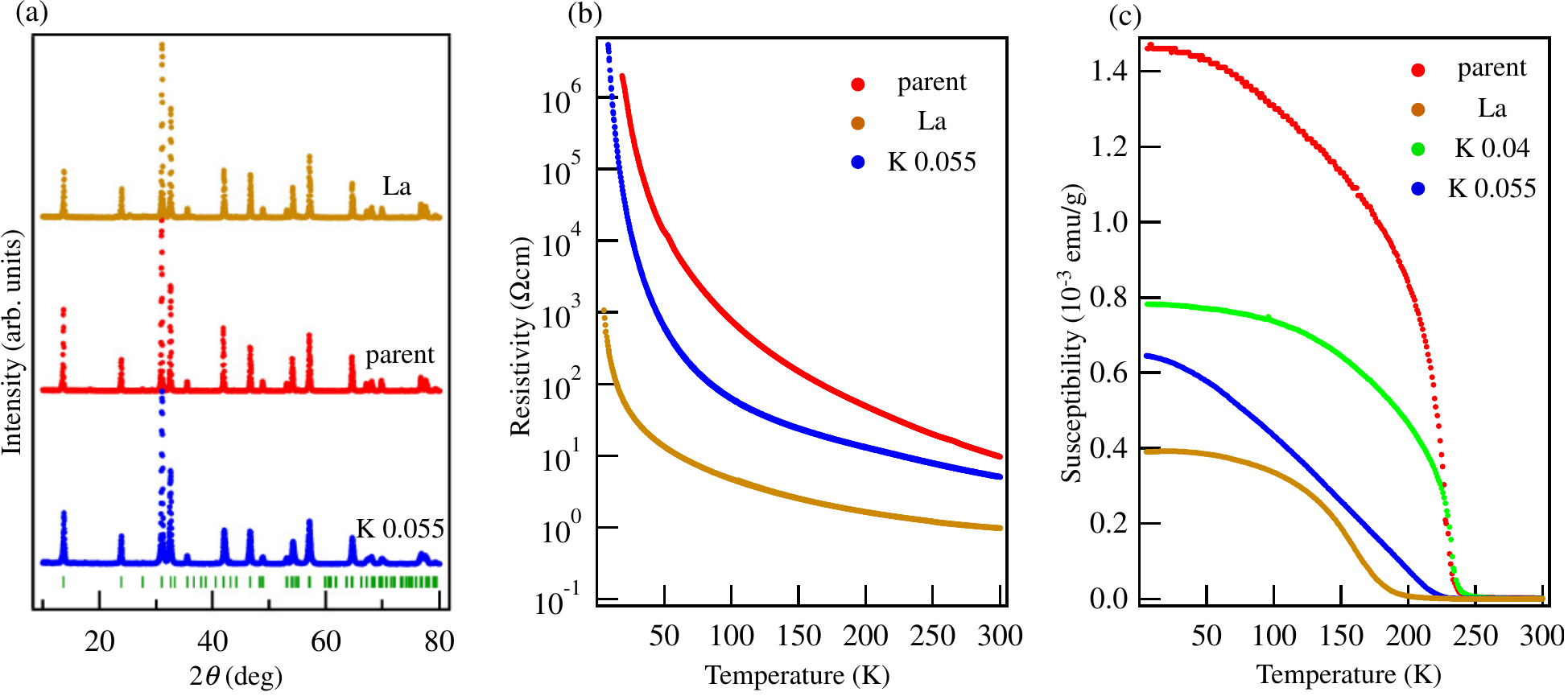}%
\caption{x-ray diffraction patterns (a) and temperature dependence of resistivity of Sr$_2$IrO$_4$, Sr$_{1.925}$La$_{0.075}$IrO$_4$, and Sr$_{1.945}$K$_{0.055}$IrO$_4$ samples. (c) Temperature dependence of the magnetic susceptibility of Sr$_2$IrO$_4$, Sr$_{1.925}$La$_{0.075}$IrO$_4$, Sr$_{1.96}$K$_{0.04}$IrO$_4$, and Sr$_{1.945}$K$_{0.055}$IrO$_4$ samples.
}
\end{figure}

\newpage

\section{Comparison of structural parameters with those of earlier reports}
	Figure S2 shows Ir-O$_p$ (a), Ir-O$_a$ (b), and Ir-Sr (c) bond distances of parent Sr$_2$IrO$_4$ taken from literature\cite{Crawford, Huang, Shimura, Ye1, Ye2, Ramarao, Bhatti, Ranjbar, Yu, Cheng1, Cheng2} including neutron diffraction, x-ray diffraction and EXAFS,  as well as those from present study.  It turned out that in the parent compound, the local structural parameters deduced in the current study show a good correspondence with the averaged structural parameters reported by neutron diffraction studies.

\begin{figure}[h]
\includegraphics[width=15cm]{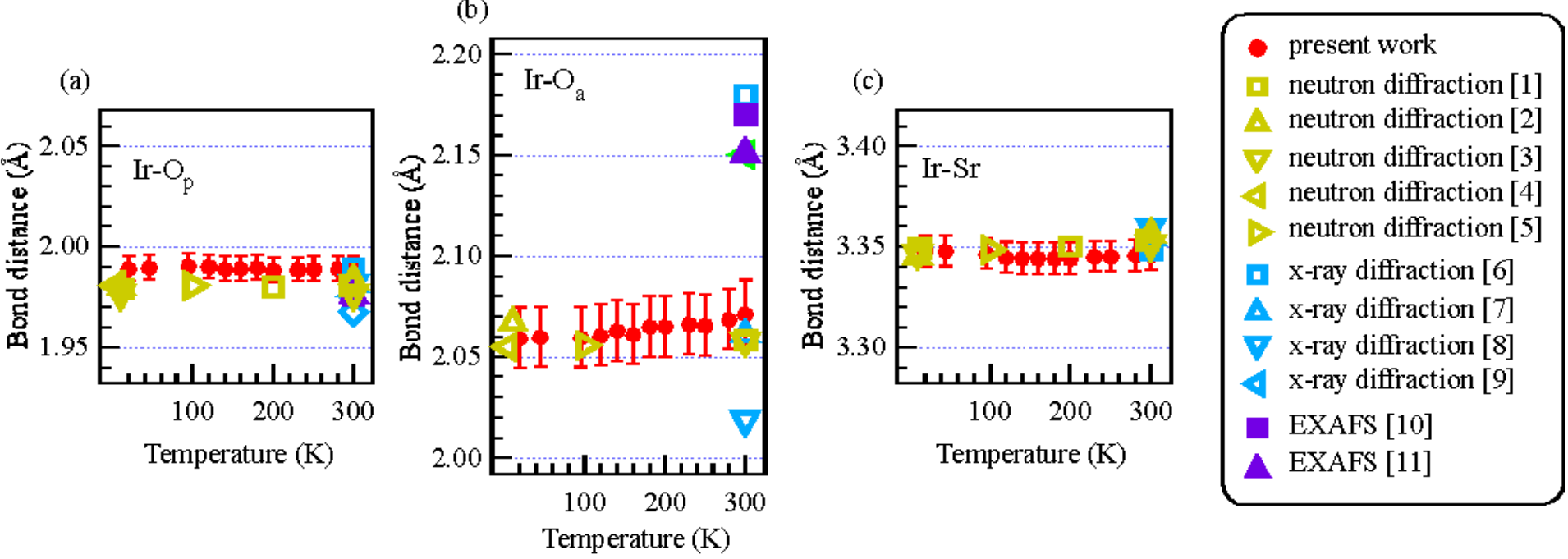}%
\caption{Comparison of Ir-O$_p$ (a), Ir-O$_a$ (b), and Ir-Sr (c) bond distances of parent Sr$_2$IrO$_4$ between the present study and earlier reports \cite{Crawford, Huang, Shimura, Ye1, Ye2, Ramarao, Bhatti, Ranjbar, Yu, Cheng1, Cheng2}.
}
\end{figure}

\end{document}